\documentclass[%
 prb, aps, bibliography, twocolumn, reprint, showpacs, nofootinbib,
 superscriptaddress, floatfix]{revtex4-2}

\usepackage{graphicx}
\usepackage{amssymb}   
\usepackage{amsfonts}
\usepackage{url}
\usepackage{amsmath}
\usepackage{color}
\usepackage{braket}
\usepackage{overpic}
\usepackage{rotating}
\usepackage{physics}
\usepackage{hyperref}
\usepackage{tikz}
\usepackage[caption=false]{subfig}
\newcommand{\be}{\begin{equation}}
\newcommand{\ee}{\end{equation}}

\begin{document}
	
\title{Topological Invariants of a Filling-Enforced Quantum Band Insulator}

\author{Abijith Krishnan}
\affiliation{Department of Physics, Massachusetts Institute of Technology, Cambridge, MA 02139, USA}
\author{Ashvin Vishwanath}
\affiliation{Department of Physics, Harvard University, Cambridge, MA 02138, USA}
\author{Hoi Chun Po}
\affiliation{Department of Physics, Massachusetts Institute of Technology, Cambridge, MA 02139, USA}

\begin{abstract}
Traditional ionic/covalent compound insulators arise from a commensuration between electron count and system volume. On the other hand, conventional topological insulators, outside of quantum hall effect systems, do not typically display such a commensuration. Tnstead, they can undergo a phase transition to a trivial insulator that preserves the electron filling. Nevertheless, in some crystalline insulators, termed filling-enforced quantum band insulators (feQBIs), electron filling can dictate nontrivial topology in the insulating ground state. Currently, little is known about the relation between feQBIs and conventional topological invariants. In this work, we study such relations for a particularly interesting example of a half-filling feQBI that is realized in space group 106 with spinless electrons. We prove that any 4-band feQBI in space group 106 with filling 2 must have a nontrivial topological invariant, namely the $\mathbb{Z}_2$ glide invariant, and thus must have a quantized magnetoelectric polarizability $\theta=\pi$. We thus have found a three-dimensional example where electron filling and band topology are tied. Such a locking raises intriguing questions about the generality of the band-inversion paradigm in describing the transition between trivial and topological phases. 

\end{abstract}
	
\maketitle

\section{Introduction}

Electrical insulators often arise from a commensuration between electron count and system volume. In traditional insulators, e.g.\ ionic/covalent compounds, electrons effectively occupy filled, localized orbitals, and the electron filling  per primitive unit cell is an integer. Alternatively, a two dimensional (2D) system of free electrons in a strong magnetic field has an insulating bulk for an integer number of electrons per magnetic flux quanta and thus exhibits the integer quantum Hall effect (IQHE).

The two described commensuration scenarios have distinct boundary electronic properties. While traditional insulators have featureless boundaries, the IQHE exemplifies the topological insulator (TI) \cite{colloquium_top_insulator,top_ins_sc, doi:10.1146/annurev-conmatphys-031214-014501}, whose bulk insulating behavior arises from quantum interference effects and can be characterized by nontrivial topological invariants, like the integer Chern number. The boundary correspondingly exhibits chiral edge modes.
Remarkably, in the IQHE, the topological invariant is tied to electron filling even in the presence of a periodic potential  \cite{Lu_Ran_Oshikawa_2020}.

Beyond the 2D electron gas, TIs exist in systems without a magnetic field, often with symmetries such as time-reversal symmetry (TRS) \cite{colloquium_top_insulator,top_ins_sc, doi:10.1146/annurev-conmatphys-031214-014501}. However, these TIs are characterized by topological invariants no longer tied to electron filling. For example, the Haldane model can be toggled between Chern number 0 and 1 upon tuning of a model parameter without changing the electron filling \cite{PhysRevLett.61.2015}. The band-inversion paradigm, which connects the phase transition between trivial and topological phases to a change in band ordering, explains this decoupling between filling and topology. Because the electron filling plays a spectator role through the topological transition, the band inversion picture implies that the topological and trivial phases have the same electron filling.

Nevertheless, in some cases, electron filling can {\it dictate} nontrivial topology in the insulating ground state of a crystal, as discussed in a series of works relating electron filling, crystalline symmetries, and band degeneracies \cite{PhysRevB.56.13607, MICHEL2001377, feqbi-soc, PhysRevLett.117.096404,  106feQBI, Bradlyn2017, PhysRevX.7.041069}. In particular,  Ref.~\onlinecite{feqbi-soc} found that band insulators can exist even when the electron filling is less than that required for forming {\it any} atomic insulator. Such insulating states do not admit a traditional localized electron description and are dubbed ``filling-enforced quantum band insulators'' (feQBIs). Symmetry-based methods for diagnosing band topology \cite{106feQBI,Bradlyn2017}, including the theory of symmetry indicators \cite{106feQBI}, have helped uncover the possible symmetry settings for realizing feQBIs.

While symmetry-based methods can reveal the presence of nontrivial band topology via symmetry representations, they do not reveal if the unveiled topological phases possess  familiar topological invariants like a Chern number or a $\mathbb Z_2$ invariant \cite{106feQBI, Bradlyn2017}.
Furthermore, the revealed phase may not be stable  to the incorporation of a trivial, atomic insulator as background degrees of freedom; such is the case for the fragile topological phases \cite{PhysRevLett.121.126402} \footnote{
A fragile TI can be trivialized by adding to it an atomic state. In contrast, conventional topological invariants like the Chern number are additive under such operation, and are never trivialized by the addition of an atomic state.
}. Recent 
 progress has helped identify the topological phases diagnosed by symmetry-based methods for some systems with TRS \cite{Song2018, PhysRevX.8.031069, PhysRevX.8.031070,Song2019}. However, the physical properties of feQBIs specifically have not yet been established. 
For example, the feQBIs discussed in Ref.~\onlinecite{feqbi-soc} are compatible with the notion of fragile topology \cite{PhysRevLett.121.126402}.
In contrast, the non-TRS feQBIs in Ref.~\onlinecite{106feQBI} are realized at electron fillings which are half-integral in units of fillings realized in atomic insulators. Such feQBIs must carry stable topological invariants. Because these feQBIs combine electron filling with stable topological invariants, they could potentially realize a true, crystalline analogue of the filling-topology correspondence in IQHE and could delimit the universality of the band inversion paradigm for crystalline TIs.

In this work, we investigate the relationship between the electron filling 
and a symmetry-protected topological invariant in a symmetry setting identified in Ref.~\onlinecite{106feQBI}. This symmetry setting admits non-TRS feQBIs at filling $2$ despite the requirement that all atomic insulators have filling in multiples of $4$. The identified invariant is the quantized magneto-electric polarizability, which manifests itself as a $\mathbb{Z}_2$ glide invariant, for Altland-Zirnbauer symmetry Class A insulators \cite{PhysRevB.55.1142} in space group 106 \cite{P3, glide_is_P3}. 
In this paper, we prove that the magnetoelectric polarizability of all $4$-band, filling $2$ feQBIs in space group 106 is necessarily $\theta = \pi \mod 2\pi$. 
We further conjecture that, analogously to the one-to-one correspondence between the Chern number and electron filling in quantum hall problems, our result implies a one-to-one correspondence between the $\theta$ angle and the filling ($4n$ vs $4n+2$) of any class A insulator in space group 106. This conjecture would imply that there is no way, at all fillings, to drive a phase transition between the two phases with distinct $\theta$ angles while respecting all specified symmetries and keeping the filling fixed. The conjectured impossibility of this transition defies the usual expectation from topological band theory, and its validity is an interesting open question.

This paper is organized as follows. In Sec.\ \ref{sec:symsetting}, we outline the necessary symmetry setting for a space group 106 feQBI with filling $2$ \cite{106feQBI}. In Sec.\ \ref{sec:model_feqbi}, we provide an example $4$-band tight-binding model of such an feQBI using gapless 2D building blocks. In Sec.\ \ref{sec:glide_invariant}, we introduce the $\mathbb{Z}_2$ glide invariant in SG 106 and in Sec.\ \ref{sec:proof}, we prove that all $4$-band, filling $2$ feQBIs in space group 106 have a nontrivial $\mathbb{Z}_2$ glide invariant \cite{glide_invariant_paper,glide_chern}. Finally, in Sec.\ \ref{sec:discussion}, we discuss how our results for the 4-band model could be extended to general feQBIs in space group 106 and elaborate on the implications of the possible extension.

\section{Overview of 4-Band Symmetry Setting} \label{sec:symsetting}
In this work, we show that any class A, half-filled insulator realized in any $4$-band model in space group (SG) 106 has the axion angle $\theta = \pi$. We first review the symmetry representations which allow for a half-filled $4$-band insulator in the SG 106 symmetry setting \cite{106feQBI}. SG 106 exists in a tetragonal Bravias lattice. Per our conventions, every unit cell has dimension $1 \times 1 \times 1$ (for simplicity, we scale the $z$-dimension to match the $x$ and $y$ dimensions). On top of lattice translations, SG 106 is generated by two symmetries: a $4_2$ screw, i.e. $(x,y,z) \mapsto (y,-x,z+1/2)$, and a vertical glide, i.e. $(x,y,z) \mapsto (1/2 + x,  1/2 - y, z)$. Thus, up to lattice translations, SG 106 has $8$ representative elements, displayed in Table \ref{tab:SG106}.

\begin{table}
\caption{\label{tab:SG106}
Representative elements of space group 106 up to lattice translations
}
\begin{tabular}{l | l}
\hline
\hline
Element   & Action on $(x,y,z)$           \\ \hline
$e$                   & $(x,y,z)$                     \\
$S$                   & $(y,-x, z + 1/2)$             \\
$\mathcal{C}_2$ & $(-x,-y,z)$                   \\
$S^3$                 & $(-y,x,z+1/2)$                \\
$X$                   & $(1/2 - x, 1/2 + y, z)$       \\
$Y$                   & $(1/2 + x, 1/2 - y, z)$       \\
$SX$                  & $(1/2 + y, 1/2 + x, 1/2 + z)$ \\
$SY$                  & $(1/2 - y, 1/2 - x, 1/2 + z)$ \\
\hline
\hline
\end{tabular}
\end{table}

\begin{figure}[t]
\begin{center}
    \includegraphics[width = 0.42 \textwidth]{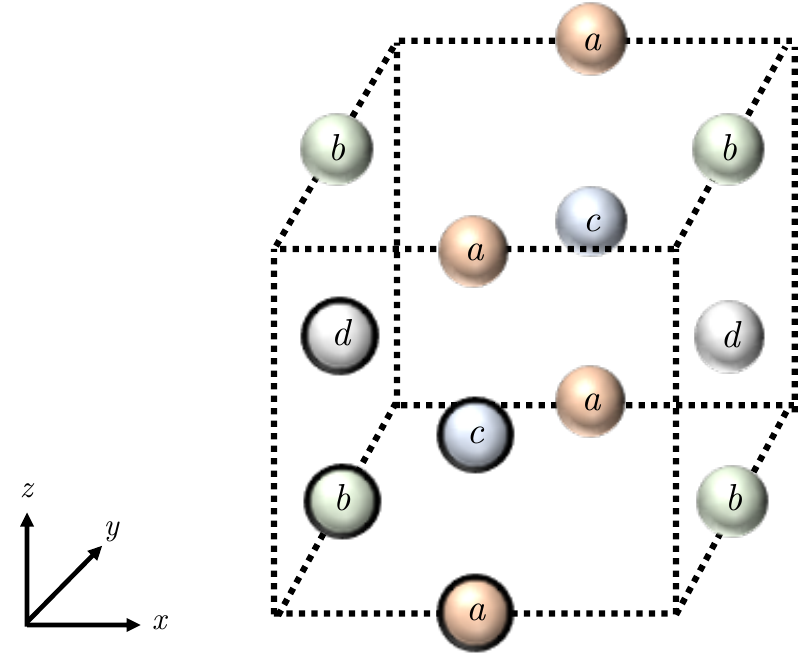}
\end{center}
\caption{A unit cell with atoms placed in Wyckoff position b in SG 106. Circles with no boundaries depict atoms in other unit cells that border the displayed unit cell.} \label{fig:wyckoffb}
\end{figure}

In SG 106, both Wyckoff positions a and b have the minimal number of sites per unit cell, which is $4$. The filling $2$ feQBIs are possible only when we consider orbitals residing in Wyckoff position b -- $(1/2,0,z_0)$, $(0,1/2,z_0)$, $(1/2,0,z_0+1/2)$, $(0,1/2,z_0 + 1/2)$ (we set $z_0=0$ for convenience) \cite{106feQBI}. Throughout, we assume spinless electrons without time-reversal symmetry, which correspond to symmetry class A. We label the annihilation operators for the orbitals at the Wyckoff sites as $a$, $b$, $c$, and $d$ respectively. A unit cell with atoms in Wyckoff position b is displayed in Fig.\ \ref{fig:wyckoffb}. (Note: the Wyckoff positions have roman letters a,b,c,..., whereas we label the sites with italic letters $a,b,c,d$.)

Moreover, orbitals in filling $2$ feQBIs in SG 106 are required to have $p$-character, i.e., the collection of orbitals has eigenvalue $-1$ under $\mathcal{C}_2 = S^2$.\footnote{More precisely, the orbitals are allowed to be any linear combination of $p_x$ and $p_y$ orbitals.} Then, for example, 
\begin{equation*}
    \mathcal{C}_2  \cdot a(1/2,0,0) = -a(-1/2,0,0).
\end{equation*} 
The degrees of freedom considered here, $p$ orbitals in Wyckoff position b, are special from the symmetry representation point of view because they support the existence of a full direct band gap at all high-symmetry momenta. More precisely, this set of orbitals leads to a pair of identical, two-dimensional irreducible representations at any of the high-symmetry points in the Brillouin zone \cite{106feQBI, Bradlyn2017}, and therefore all compatibility relations can be satisfied with only half of the bands. Further model calculations show that the gap can be sustained everywhere in the Brillouin zone, and therefore a band insulator is attainable at filling 2 \cite{106feQBI}.

Alternatively, we could have put $s$ orbitals at each site or considered sites in Wyckoff position a. The latter case is distinct from Wyckoff position b because the $X$ and $Y$ symmetries would send each orbital to itself, up to lattice translations and a minus sign. Either or both of these choices would violate the compatibility relations at half filling, and a feQBI is thus impossible with those degrees of freedom.

Next, we deduce the action of SG 106 elements on the Wyckoff sites. For simplicity, we perform this analysis up to Bravais lattice translations, as Bravais lattice translations do not factor into our later analysis \footnote{Orbitals related by a lattice translation have identical transformation properties, and so for the analysis it suffices the relate the symmetry elements up to translations.}. First, notice that $S^2 = \mathcal{C}_2$ and that $XY = \mathcal{C}_2$.  Without loss of generality (the phase is either $\pi/2$ or $-\pi/2$), we assign
\begin{align*}
    S  \cdot a(1/2,0,0) = i d(0,-1/2,1/2).
\end{align*}
Additionally, using the $p$-character of the orbitals, we assign, without loss of generality (either $X$ or $Y$ must have a negative sign in its definition)
\begin{align*}
    X \cdot a(1/2,0,0) = b(0,1/2,0),\\  Y \cdot a(1/2,0,0) = -b(1,1/2,0).
\end{align*}
Using the relation $SXS^{-1} = Y$ (which holds up to Bravais lattice translation), we generate the action of all other elements on the electronic orbitals, as shown in Table \ref{tab:action}. With these real-space symmetry relations, we also generate the momentum-space symmetry matrices as follows. We order the orbitals as $(\mathcal{O}^i) = (a, b, c, d)$, and for each symmetry $G$, we compute $M_G^{ij}(\vb{k})$, where $G(\vb{k}) = \mathcal{O}^i(G \vb{k}) \cdot {M_G}^{ij}(\vb{k}) \cdot \mathcal{O}^j(\vb{k})^\dagger$. The symmetry matrices are thus given by
\begin{align}
    M_X(\vb{k}) &= \begin{pmatrix}0 & e^{-ik_y} & 0 & 0\\ 
1 & 0 & 0 & 0\\
0 & 0 & 0 & -e^{-ik_y}\\
0 & 0 & -1 & 0
\end{pmatrix}, \nonumber \\
M_Y(\vb{k}) &= \begin{pmatrix}0 & -1 & 0 & 0\\
-e^{-ik_x} & 0 & 0 & 0\\
0 & 0 & 0 & 1\\
0 & 0 & e^{-ik_x} & 0
\end{pmatrix}, \nonumber \\
M_S(\vb{k}) &= i\begin{pmatrix}
0 & 0 & 0 & e^{-ik_z}\\
0 & 0 & e^{-ik_x - ik_z} & 0\\
0 & 1 & 0 & 0\\
e^{-ik_x} & 0 & 0 & 0
\end{pmatrix}.
\end{align}
We obtain these matrices using the convention
$$\mathcal{O}^j(\vb{k})^\dagger = \frac{1}{\sqrt{N}}\sum_{\vb{r}} e^{i\vb{k} \cdot \vb{r}}\mathcal{O}^j(\vb{r})^\dagger,$$
where $N$ is the number of lattice sites, and $\vb{r}$ is the location of each Bravais lattice unit cell.
The corresponding symmetry relations in momentum space are given as follows. For symmetry elements $G$, $H$, and $G'$, and a lattice translation $L(\vb{r})$, suppose the identity $GH = L(\vb{r})G'$ holds in position space. Then, in momentum space, the identity takes the form 
\begin{equation}\label{eq:sym_eq}
    G(H\vb{k}) H(\vb{k}) = e^{-i\vb{r} \cdot G'(\vb{k})} G'(\vb{k}).
\end{equation}

\begin{table}[b]
\caption{\label{tab:action}%
Action of each representative element of SG 106, $G$, from Table \ref{tab:SG106}, on four representative electronic orbitals $a_0$, $b_0$, $c_0$, $d_0$. We use the shorthand $O_{0}$ to refer to the orbital $O$ with lower right bottom corner at the origin, and $O_{l,m,n}$ to refer to the orbital $O$ with lower right bottom corner at the origin $(l,m,n)$.  
}
\begin{tabular}{ l|l|l|l|l } 
 \hline
\hline
$G$ & $G \cdot a_0$ & $G \cdot b_0$ & $G \cdot c_0$ & $G \cdot d_0$  \\ \hline
 $e$ & $a_0$ & $b_0$ & $c_0$ & $d_0$ \\ \hline
 $S$ & $id_{0,-1,0}$ & $ic_0$ & $ib_{0,-1,1} $ & $ia_{0,0,1}$ \\ \hline
 $\mathcal{C}_2$ & $-a_{-1,0,0}$ & $-b_{0,-1,0}$ & $-c_{-1,0,0}$ & $-d_{0,-1,0}$ \\ \hline
 $S^3$ & $-id_{0}$ & $-ic_{-1,0,0}$ &$-ib_{0,0,1}$ &$-ia_{-1,0,1}$ \\ \hline
 $X$ & $b_0$ & $a_{0,1,0}$ & $-d_0$ & $-c_{0,1,0}$ \\ \hline
 $Y$ & $-b_{1,0,0}$ & $-a_0$ & $d_{1,0,0}$ & $c_0$ \\ \hline
 $SX$ & $ic_{0,1,0}$ & $id_{1,0,0}$ & $-ia_{0,1,1}$ &$-ib_{1,0,1}$ \\ \hline
 $SY$ & $-ic_0$ & $-id_0$ & $ ia_{0,0,1}$& $ib_{0,0,1}$ \\
\hline
\hline
\end{tabular}
\end{table}

\section{2D Building Blocks of 4-Band feQBIs} \label{sec:model_feqbi}
\begin{figure}[ttp]
\begin{center}
\begin{tikzpicture}[scale = 1.5]

\draw[->] (-1.5, 0) -- (-1, 0);
\draw[->] (-1.5, 0) -- (-1.5, 0.5);

\node at (-0.8, 0) {$x$};
\node at (-1.5, 0.7) {$y$};

\draw[thick] (-1.8, 2.3) -- (-1.2, 2.3);
\filldraw[gray] (-1.5, 2.3) circle (2pt);
\node at (-0.9, 2.3) {$a/c$};
\draw[ultra thick, dashed] (-1.8, 2) -- (-1.2, 2);
\filldraw[gray!40] (-1.5, 2) circle (2pt);
\node at (-0.9, 2) {$b/d$};

\draw (-2, 2.6) -- (-0.6, 2.6) -- (-0.6, 1.7) -- (-2,1.7) -- (-2,2.6);

\draw[ultra thick, dashed](0.5, 0) -- (3, 2.5);
\draw[ultra thick, dashed](0.0, 0.5) -- (2.5, 3);
\draw[ultra thick, dashed](1.5, 0) -- (3, 1.5);
\draw[ultra thick, dashed](0, 1.5) -- (1.5, 3);

\draw[thick](2.5, 0) -- (0, 2.5);
\draw[thick](3, 0.5) -- (0.5, 3);
\draw[thick](0, 1.5) -- (1.5, 0.0);
\draw[thick](3, 1.5) -- (1.5, 3);
\filldraw[gray] (0.5, 0) circle (2pt);
\filldraw[gray] (1.5, 0) circle (2pt);
\filldraw[gray] (2.5, 0) circle (2pt);
\filldraw[gray] (1.5, 1) circle (2pt);
\filldraw[gray] (2.5, 1) circle (2pt);
\filldraw[gray] (0.5, 1) circle (2pt);
\filldraw[gray] (1.5, 2) circle (2pt);
\filldraw[gray] (2.5, 2) circle (2pt);
\filldraw[gray] (0.5, 2) circle (2pt);
\filldraw[gray] (1.5, 3) circle (2pt);
\filldraw[gray] (2.5, 3) circle (2pt);
\filldraw[gray] (0.5, 3) circle (2pt);

\draw[ultra thick, gray] (1.5,3) arc (90:0:1);
\draw[ultra thick, gray] (1.5,3) arc (225:585:0.2);

\filldraw[gray!40] (0, 0.5) circle (2pt);
\filldraw[gray!40] (0, 1.5) circle (2pt);
\filldraw[gray!40] (0, 2.5) circle (2pt);
\filldraw[gray!40] (1, 0.5) circle (2pt);
\filldraw[gray!40] (1, 1.5) circle (2pt);
\filldraw[gray!40] (1, 2.5) circle (2pt);
\filldraw[gray!40] (2, 0.5) circle (2pt);
\filldraw[gray!40] (2, 1.5) circle (2pt);
\filldraw[gray!40] (2, 2.5) circle (2pt);
\filldraw[gray!40] (3, 0.5) circle (2pt);
\filldraw[gray!40] (3, 1.5) circle (2pt);
\filldraw[gray!40] (3, 2.5) circle (2pt);

\draw[ultra thick, gray!40] (1, 1.5) arc (180:90:1);
\draw[ultra thick, gray!40] (1, 1.5) arc (225:585:0.2);

\draw[densely dotted, thick](1.9, 0.9) -- (0.9, 0.9) -- (0.9, 1.9) -- (1.9, 1.9) -- (1.9, 0.9);

\end{tikzpicture}
\end{center}
\caption{Top down view of a SG 106 crystal with the $a/c$ orbitals in dark gray, and the $b/d$ orbitals in light gray. The crystal can be described as a set of gapless 2D planes of coupled $a$ and $c$ orbitals (solid line) and an intersecting set of gapless 2D planes of $b$ and $d$ orbitals (dashed line) that are coupled together. The shortest allowed bonds are drawn for one of the $a-c$ planes and one of the $b-d$ planes. A unit cell of the crystal is depicted via a dotted square.} \label{fig:gaplessbuildingblocks}
\end{figure}
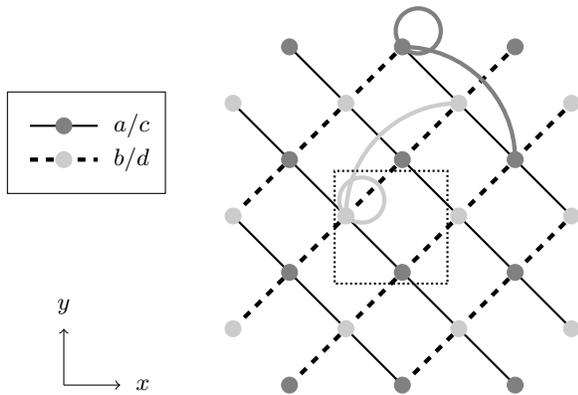

Now that we have described the necessary symmetry setting for a 4-band, filling $2$ feQBI, we construct an example model of the feQBI from which we can glean insight. We emphasize that our proof in the next sections is not specifically tied with the model construction we present here.
Topological crystalline insulators in 3D are typically studied by analyzing the properties of lower-dimensional analogues that respect the symmetries of the 3D insulator, i.e., 1D and 2D electronic phases occupying suitable subregions of the 3D crystals. In this section, we take a similar approach for our SG 106 symmetry setting by first considering $2$-band systems on the intersecting planes given by 
\begin{equation} \label{eq:planes}
    x + y = (n + 1/2), \quad x - y = (n + 1/2),
\end{equation}
for integers $n$. Because they respect the diagonal glide symmetries, $SX$ and $SY$, up to lattice translations, these planes are natural choices for lower-dimensional analogues. We populate the former planes in Eq.\ \eqref{eq:planes} with $a$ and $c$ orbitals and the latter planes with $b$ and $d$ orbitals. In other words, the only allowed bonds are along the $x + y = (n+1/2)$ planes between $a$ and $c$ orbitals and along the $x-y = (n+1/2)$ planes between $b$ and $d$ orbitals. The $a-c$ and $b-d$ planes are related by the $X$ symmetry. A schematic of these intersecting planes are shown in Fig.\ \ref{fig:gaplessbuildingblocks}.

The SG 106 symmetries preserve the intersecting planes as follows. The symmetries equivalent to $SX$, $SY$, $C_2$, and $e$ up to translation (see Table \ref{tab:SG106}) map the $a-c$ and $b-d$ planes to other $a-c$ and $b-d$ planes respectively while the other symmetries map the $a-c$ and $b-d$ planes to each other. In fact, after application of symmetry $SX$ and a lattice translation of $(-1,0,0)$, each $a-c$ plane is mapped to itself. Likewise, after application of symmetry $SY$, each $b-d$ plane is mapped to itself. Thus, on these decoupled glide-symmetric planes, the SG 106 symmetries effectively act as 2D non-symmorphic symmetries on each plane.

Generally, non-symmorphic symmetries increase the minimum filling at which insulators are possible \cite{MICHEL2001377, Konig3502, PhysRevB.94.195109}, and any $2$-band system that solely lives on these glide-symmetric planes will always be gapless. Let us revisit the argument for our particular symmetry setting. Because the $b-d$ planes are related to the $a-c$ planes by the $X$ glide, we restrict our attention to the $a-c$ planes. We focus on the symmetry given by application of $SX$ and a lattice translation of $(-1,0,0)$, as this maps each $a-c$ plane to itself. We compute the symmetry matrix with Eq. \eqref{eq:sym_eq} as $e^{ik_x}M_S(X\vb{k})M_X(\vb{k})$:
$$ \begin{pmatrix}
0 & 0 & -e^{ik_x-ik_z}&0 \\
0 & 0 & 0 & -e^{2ik_x - ik_y - ik_z} \\ 
e^{ik_x} & 0 & 0 &0 \\
0& e^{2ik_x- i k_y} & 0 & 0 
\end{pmatrix}.$$
As expected from our real-space analysis of the symmetries (that $SX$ preserves the set of $a-c$ planes and $b-d$ planes), this symmetry matrix acts on the $a-c$ planes and $b-d$ planes separately. We are thus free to restrict our attention to the action of this symmetry solely on the $a-c$ planes:
$$\begin{pmatrix}
0 & -e^{ik_x - k_z}\\
e^{ik_x} & 0
\end{pmatrix}.$$

Under this symmetry, a 2D Hamiltonian on the a-c planes takes the form
\begin{equation}
    H_{a-c}(\vb{k}) =E_{a-c}(\vb{k}) \mathbb{I} + r_{a-c}(\vb{k})\begin{pmatrix}0 & i e^{-ik_z/2}\\-ie^{ik_z/2} & 0
\end{pmatrix}
\end{equation}
where $E_{a-c}$ and $r_{a-c}$ are both real functions. Because the Hamiltonian is invariant under $k_z \to k_z + 2\pi$, we get the condition 
$r_{a-c}(\vb{k}) = -r_{a-c}(\vb{k} + 2\pi \vu{z})$. Thus, $r_{a-c}$ must have a $0$, and $H_{a-c}$ is gapless. 

Because each set of intersecting plane is gapless, and the sets are related to each other via the $X$ glide, the above argument indicates that constructing a feQBI with these 2-D decoupled intersecting planes is impossible. We can extend this argument to illustrate that any band insulator in this symmetry setting realized at the filling of $2$ cannot be atomic. If there were a filling $2$ atomic insulator, its Hamiltonian could be deformed into one whose ground state is a product state of two filled orbitals. However, because the glide symmetry forces us to assign one orbital per unit cell to each set of intersecting planes, which we just showed to be gapless, no atomic state can exist at filling $2$.

\begin{figure}[ttp]
    \centering
    \includegraphics[width = 0.48\textwidth]{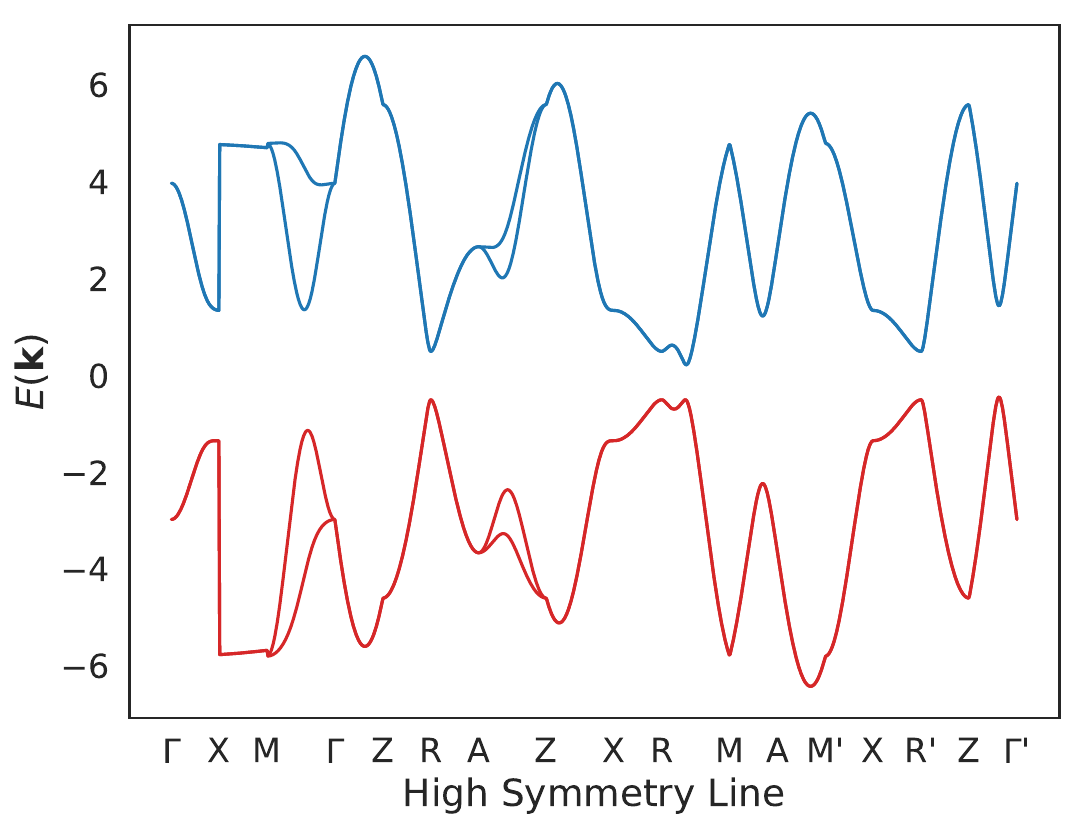}
    \caption{Band structure eigenvalues $E(\vb{k})$ for a SG 106 feQBI with bonds $t_x = 2$, $t_z = 2 + i$, $t_c = 2 + 3i$, and $t_d = 12- 20i$, and initial Hamiltonian given in Eq.\ \eqref{eq:ansatz}.}
    \label{fig:band_structure}
\end{figure}

Nevertheless, we can build an feQBI by coupling the 2D intersecting planes. We define the function $P_G$, which, for an order $2$ group symmetry $G$ and Hamiltonian $H$, returns a new Hamiltonian symmetric under $G$:
$$P_G \circ H = \frac{1}{2}\left[H + M_G(G^{-1}\vb{k})H(G^{-1}\vb{k})M_G(G^{-1}\vb{k})^{-1}\right].$$
Then, because $X$ and $Y$ are both order $2$ symmetries, and because $S^2$ maps the subgroup spanned by $X$ and $Y$ to itself, we can generate a SG 106 Hamiltonian $H_{106}$ from any tight-binding Hamiltonian $H_0$ by computing 
\begin{equation}\label{eq:sym_ansatz}
H_{106}(\vb{k}) = P_S \circ P_Y \circ P_X \circ H_0(\vb{k}).
\end{equation}
In our case, we use the following $H_0(\vb{k})$:
\begin{equation}\label{eq:ansatz}
    \begin{pmatrix} 2 t_x \cos(k_x) & 0 & t_z + t_c \cos(k_x - k_y) & t_d\\ 
0 & 0 & 0 & 0\\
t_z^\star + t_c^\star \cos(k_x - k_y) & 0 & 0 & 0\\
t_d^\star & 0 & 0 & 0
\end{pmatrix}
\end{equation}
Here, $t_z$ and $t_c$ are the strengths of bonds on the intersecting planes, $t_x$ is a nonzero nonchiral coupling between the $ac$ and between the $bd$ planes, and $t_d$ is a sufficiently strong bond between the intersecting planes. All bonds are complex except for $t_x$. Fig.\ \ref{fig:band_structure} displays the band structure of a feQBI with initial Hamiltonian given by Eq.\ \eqref{eq:ansatz} for bonds $t_x = 2$, $t_z = 2 + i$, $t_c = 2 + 3i$, and $t_d = 12- 20i$. Eq.\ \eqref{eq:full_model_hamiltonian} in Appendix \ref{appendix-full-hamiltonian} gives the full form of $H_{106}(\vb{k})$.

\section{Glide Invariant in Space Group 106} \label{sec:glide_invariant}

We now characterize the nontrivial topology of $4$-band feQBIs in SG 106. A natural choice of topological invariant is the $\mathbb{Z}_2$ glide invariant, which is introduced in Ref.~\onlinecite{glide_invariant_paper, glide_chern}, and is equivalent to the more general magnetoelectric polarizability, or $\theta$-angle \cite{P3, glide_is_P3}.  We first give a brief description of the glide invariant for the symmetry $X$ before proving that all feQBIs in the SG 106 symmetry setting have the glide invariant. 

To compute our glide invariant, we first simultaneously diagonalize our SG 106 Hamiltonian $H_{106}(\vb{k})$ and the symmetry matrix $M_X(\vb{k})$ on the glide invariant planes in $\vb{k}$-space: $k_x = 0$ and $k_x = \pi$ \cite{glide_invariant_paper}. Suppose our SG 106 Hamiltonian has filling $2n$. Because $M_X(\vb{k})$ has two eigenvalues, $\pm e^{-ik_y/2}$, each of the filled states belong to the ``positive," or $+e^{-ik_y/2}$, sector or the ''negative," or $-e^{-ik_y/2}$, sector. We thus label the filled states of $H_{106}$ on the glide invariant planes as $v_{i\pm}(k_y, k_z)$, where $i$ indexes the $n$ eigenvectors in each sector. Because $\pm e^{-i k_y/2} \mapsto \mp e^{-ik_y/2}$ under $2\pi$ translations in $k_y$, the eigenstate $v_{i\pm}$ have the following periodicities:
\be \label{eq:periodicity}
v_{i\pm}(k_y, k_z) = v_{i\pm}(k_y, k_z + 2\pi) = v_{i \mp}(k_y + 2\pi, k_z).
\ee

\begin{figure*}
    \centering
    \subfloat[]{\includegraphics[height= 0.34\textwidth]{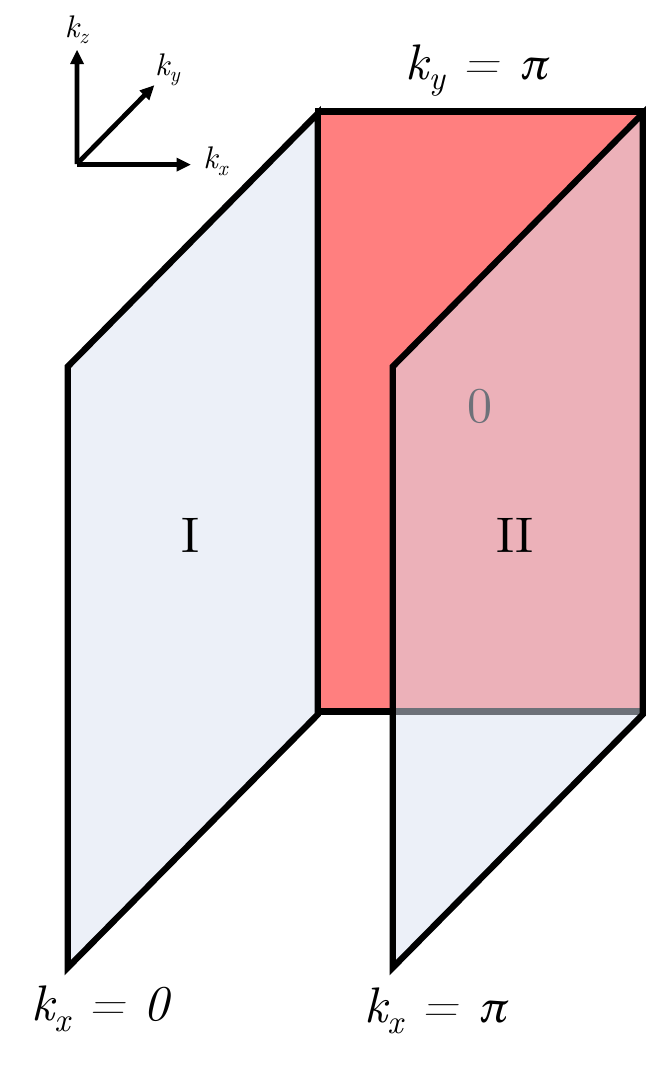}}
    \hfill
    \subfloat[]{\includegraphics[height= 0.34\textwidth]{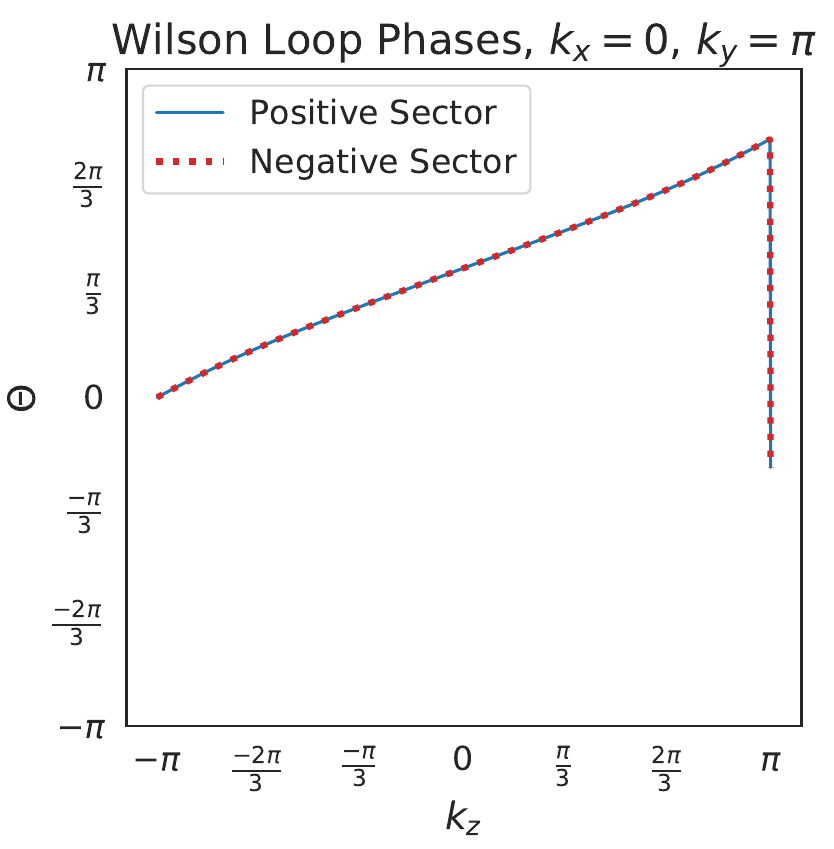}}
    \hfill
    \subfloat[]{\includegraphics[height= 0.34\textwidth]{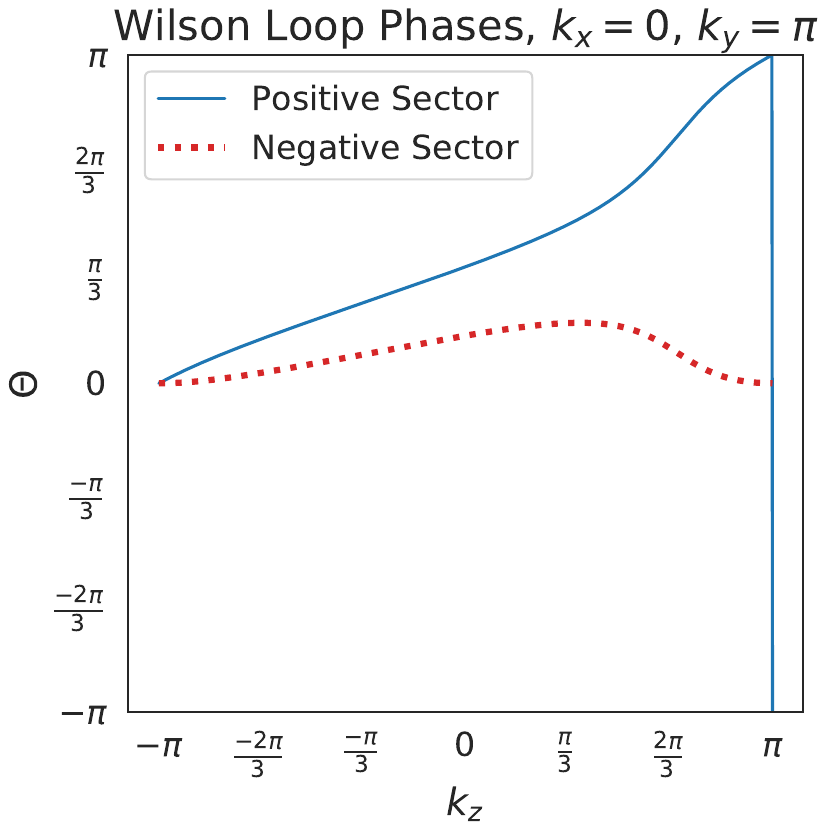}} \\
    \subfloat[]{\includegraphics[height= 0.29\textwidth]{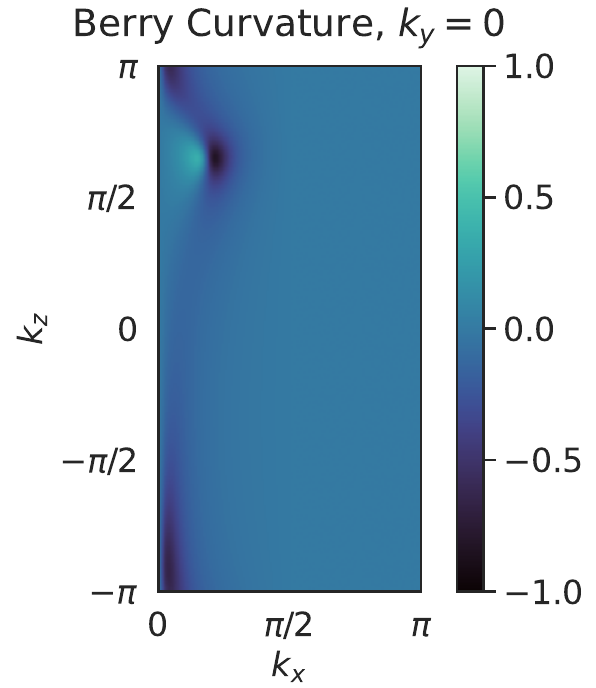}}
    \hfill
    \subfloat[]{\includegraphics[height= 0.29\textwidth]{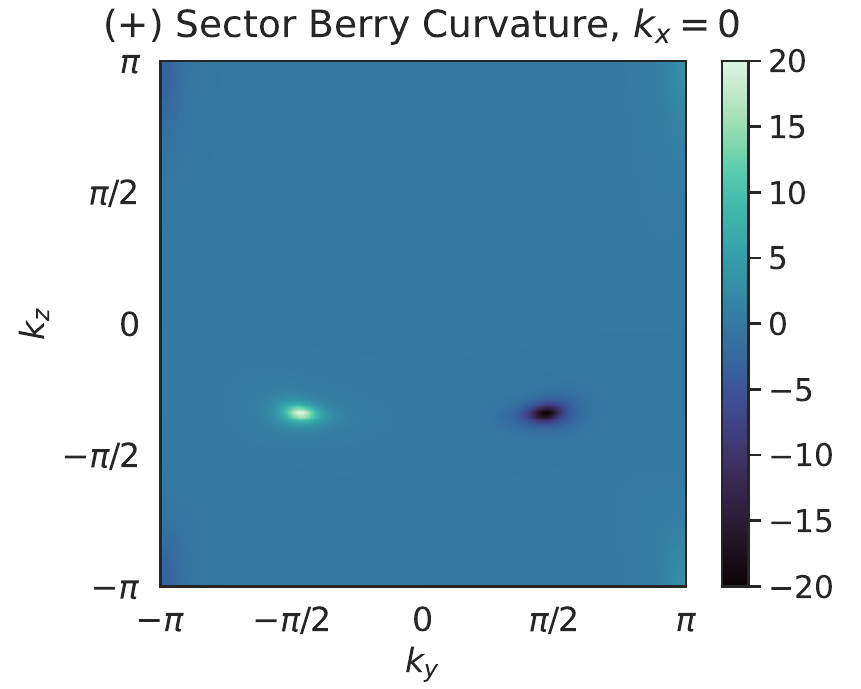}}
    \hfill
    \subfloat[]{\includegraphics[height= 0.29\textwidth]{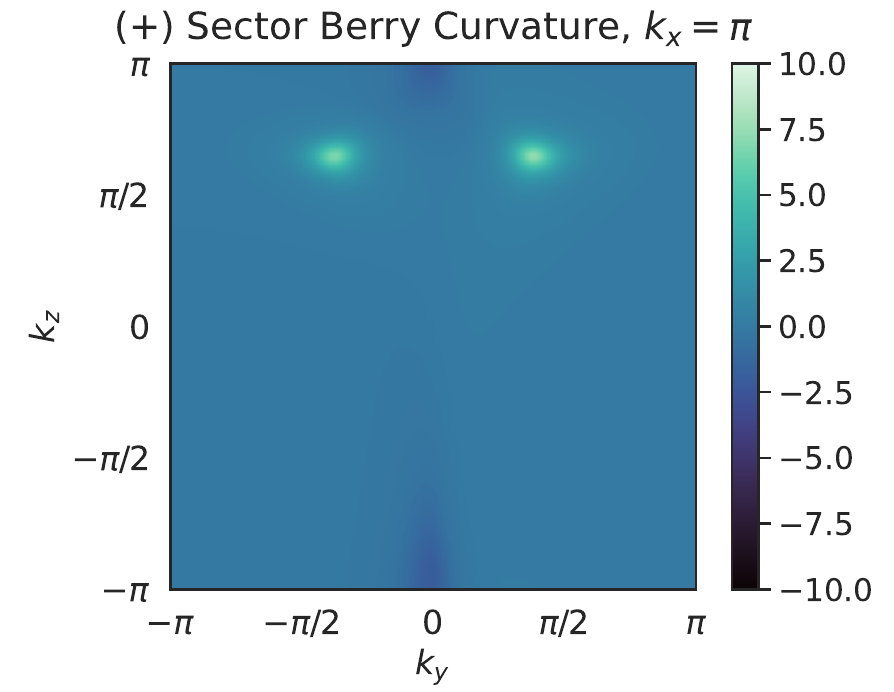}} 
    \caption{Numerical calculations of the glide invariant for the model in Equations \eqref{eq:sym_ansatz}, \eqref{eq:ansatz}. (a) Schematic of the glide invariant planes I and II and the half-plane 0. (b,c) Berry phase integrals $\Theta(k_z)$ along the $k_y = \pi$ line for plane I (e) and plane II (f). (d-f) Computation of the positive sector Berry curvature on planes I and II (e and f) and the total Berry curvature on half-plane 0 (d).}
    \label{fig:glide_calculation}
\end{figure*}

The glide invariant is given by the sum of three integers, $n_0$, $n_I$, and $n_{II}$, modulo $2$, each associated with a different Brillouin zone plane \cite{glide_invariant_paper}. Planes I and II are the glide invariant planes $k_x=0$ and $k_x = \pi$ respectively, and half-plane 0 is the $k_y = \pi$, $0 \le k_x \le \pi$ half-plane. Because of the other symmetries in SG 106, there is no net Chern number on planes I and II (see Appendix \ref{appendix-chern}). Planes I and II and half-plane 0 are depicted in Fig.\ \ref{fig:glide_calculation}(a). 

We first subtract the integral of the positive sector Berry connection on a loop around plane I from the integral of the positive sector Berry curvature on plane I. This number is always a multiple of $2\pi$, so we define 
\begin{align}\label{eq:n1}
    2\pi n_I = \int_{I} \tr F^{I}_+ - \int_{-\pi}^\pi \dd k_z \tr (\vb{A}^I_+)^z (\pi, k_z) \nonumber\\
    + \int_{-\pi}^\pi \dd k_z \tr (\vb{A}^I_+)^z (-\pi, k_z).
\end{align}
Here we use that the contributions from the line segments $k_x = 0$, $k_z = \pm \pi$ cancel each other, but the contributions from $k_x = 0$, $k_y = \pm \pi$ do not as a consequence of Eq.\ \eqref{eq:periodicity}. We compute $n_{II}$ similarly: 
\begin{align}
    2\pi n_{II} = \int_{II} \tr F^{I}_+ - \int_{-\pi}^\pi \dd k_z \tr (\vb{A}^{II}_+)^z (\pi, k_z) \nonumber\\
    + \int_{-\pi}^\pi \dd k_z \tr (\vb{A}^{II}_+)^z (-\pi, k_z).
\end{align}
Finally, we subtract the integral of the total Berry connection on a loop around half-plane 0 from the integral of the total Berry curvature on half-plane 0. Then,
\begin{align}
    2\pi n_{0} = &\int_{0} \tr F - \int_{-\pi}^\pi \dd k_z \big[\tr (\vb{A}^{II}_+)^z (\pi, k_z) + \nonumber\\ &\tr (\vb{A}^{II}_+)^z (-\pi, k_z)\big] 
    + \int_{-\pi}^\pi \dd k_z \big[\tr (\vb{A}^{I}_+)^z (\pi, k_z) + \nonumber\\ &\tr (\vb{A}^{I}_+)^z (-\pi, k_z)\big].
\end{align}
Here, we use that the contributions from the line segments $k_y = \pi$, $k_z = \pm \pi$ to the Berry connection integral around half-plane 0 cancel each other. We also use that  $(\vb{A}^{I/II}_+)(k_y, -\pi) = (\vb{A}^{I/II}_-)(k_y, \pi)$, a consequence of Eq.\ \eqref{eq:periodicity}. Finally, the glide invariant is given by
\be
n = n_0 + n_{I} + n_{II} \mod 2.
\ee
In Fig.\ \ref{fig:glide_calculation}, we display the configuration of the glide-invariant planes, and we compute both the Berry phase integrals
$$\Theta(k_y = \pi, k_z)^{I/II}_{\pm} = \int_{-\pi}^{k_z} \tr(\vb{A}^{I/II}_\pm)^z(k_y, k_z') \dd k_z'$$
and the Berry curvatures
$$\tr F_+^{I}(k_y, k_z), \quad \tr F_+^{II}(k_y, k_z), \quad  \tr F(k_x, k_z)$$
for the model in Eqs.\ \eqref{eq:sym_ansatz} and \eqref{eq:ansatz}. Our calculations result in a nontrivial glide invariant ($n_0 = 0$, $n_{I} = 0$, and $n_{II} = 1$, though these individual integers may vary based on choice of gauge).

\section{Glide Invariant for 4-band SG 106 feQBIs} \label{sec:proof}
We now prove that in the necessary symmetry setting (Wyckoff position b, $p$-character orbitals, spinless electrons -- see Sec.\ \ref{sec:symsetting}) for 4-band, half-filled insulators in SG 106, the $Y$ glide and $S$ screw symmetries mandate that the glide invariant be nontrivial. In our derivation, we find it convenient to work in the $X$ glide eigenbasis, given by the columns of the below matrix: 
\be\label{xbasis}
B_X(\vb{k}) = \frac{1}{\sqrt{2}}\begin{pmatrix}e^{-ik_y/2} & 0 & -e^{-ik_y/2} & 0 \\
1 & 0 & 1 & 0 \\
0 & -e^{-ik_y/2} & 0 & e^{-ik_y/2}\\
0 & 1 & 0 & 1
\end{pmatrix}.
\ee
The first two columns of $B_X$ span the positive ($+e^{ik_y/2}$ eigenvalue) sector while the last two columns span the negative ($-e^{ik_y/2}$ eigenvalue) sector. Note that $B_X$ is periodic under $k_y \to k_y + 4\pi$, and the first two columsn swap with the last two columns under $k_y \to k_y + 2\pi$.

This choice of basis block-diagonalizes the SG 106 Hamiltonian on the glide-invariant planes. Moreover, because the $k_y \to k_y + 2\pi$ symmetry swaps the columns of $B_X$, the Hamiltonian in the $X$ glide eigenbasis is 
\be \label{block-diagonal-hamiltonian}
H^{p}(k_y, k_z) = \begin{pmatrix}
h_\mu^{p}(k_y, k_z) \sigma^\mu & 0 \\
0 & h_\mu^{p}(k_y + 2\pi, k_z) \sigma^\mu
\end{pmatrix},
\ee
where $p$ is an index labeling the glide-invariant plane ($I$/$II$), $\sigma^\mu$ is the Pauli matrix basis, and $h_\mu^p(k_y, k_z)$ is a $4$-tuplet of functions periodic under $k_y \to k_y + 4\pi$, $k_z \to k_z + 2\pi$. We compute $H^p$ from the SG 106 Hamiltonian $H_{106}$ as follows:
$$H^p(k_y, k_z) = B_X(\vb{k})^\dagger H_{106}(\vb{k})B_X(\vb{k}),$$
where the $x$-component of $\vb{k}$ corresponds to plane $p$. If $H_{106}$ describes an feQBI, each block of $H^p$ must be gapped, i.e., $(h_1^p, h_2^p, h_3^p) = \vec{h}^p$ must have nonzero magnitude.

On the glide invariant planes I and II, we restrict $H^p$ further using the $Y$-glide symmetry. On Plane I, the $Y$ glide takes the form
\be
M_Y^I(\vb{k}) = B_X(Y\vb{k})^\dagger M_Y(\vb{k}) B_X(\vb{k}) = e^{-ik_y/2} \begin{pmatrix}-\mathbb{I} & 0 \\ 0 & \mathbb{I}\end{pmatrix}, \nonumber
\ee
where $\mathbb{I} = \sigma^0$ is the $2\times 2$ identity matrix, and the $x$-component of $\vb{k}$ is $0$. Under the $Y$ symmetry, the functions $h^I_\mu$ satisfy 
\be \label{eq:ysym-plane1}
h_\mu^I(k_y, k_z) = h_\mu^I(-k_y, k_z).
\ee
Likewise, on Plane II, the $Y$ glide takes the form
\be
M_Y^{II}(\vb{k}) = B_X(Y\vb{k})^\dagger M_Y(\vb{k}) B_X(\vb{k}) = e^{-ik_y/2} \begin{pmatrix}0& -\mathbb{I}  \\ \mathbb{I} & 0\end{pmatrix}, \nonumber
\ee
where the $x$-component of $\vb{k}$ is $\pi$. Under the $Y$ symmetry, the functions $h^{II}_\mu$ satisfy 
\be \label{eq:ysym-plane0}
h_\mu^{II}(k_y, k_z) = h_\mu^{II}(2\pi-k_y, k_z).
\ee

Finally, on half-plane 0, we work in the $Y$ glide eigenbasis, given by the columns of the below matrix:
$$B_Y(\vb{k}) = \frac{1}{\sqrt{2}}
\begin{pmatrix}
0 & 1  & 0 & 1\\
0 & -e^{-ik_x/2} & 0 &  e^{-ik_x/2} \\
 1 & 0 & 1 & 0\\
 e^{-ik_x/2} & 0 & -e^{-ik_x/2} & 0
\end{pmatrix}.$$
In this basis, the Hamiltonian is block diagonal on half-plane 0:
\be \label{}
H^{0}(k_x, k_z) = \begin{pmatrix}
h_\mu^{0}(k_x, k_z) \sigma^\mu & 0 \\
0 & h_\mu^{0}(k_x + 2\pi, k_z) \sigma^\mu
\end{pmatrix}.
\ee

During our derivation, we sometimes for convenience will denote the Berry phase integral as $\lambda_{\pm}^{I/II}(k_y)$, where 
\be
\lambda^{I/II}_{\pm}(k_y) = \int_{-\pi}^\pi \tr (\vb{A}^{I/II}_{\pm})^z(k_y, k_z) \dd k_z.
\ee

\subsection{Plane I Contribution}
We now compute $n_I$, the integer associated with Plane I, by constraining the Hamiltonian $H^I$ using the $Y$ glide. 

As Eq.\ \eqref{eq:n1} states, the integer $n_I$ satisfies
\begin{align*}
    2\pi n_I = \int_{I} \tr F^{I}_+ - \lambda^I_+(\pi) + \lambda^I_-(-\pi).
\end{align*}
Even though the basis of $H^I$ is $\vb{k}$-dependent, we directly use the positive sector eigenstates of $H^I$ to compute the Berry connection and curvature contributions to $n_I$ (see Appendix \ref{appendix-glide} for a proof). Because $H^p$ restricted to the positive sector is just the $2\times 2$ matrix $h_\mu(k_y, k_z) \sigma^\mu$, we use the standard results for the Berry connection and curvature for a two-level system \cite{Carpentier2017}:
\begin{align}
    \vb{A}^I_+ &= \frac{h_1^I \grad_{\vb{k}}h_2^I - h_2^I \grad_{\vb{k}}h_1^I}{2|\vec{h}^I|(\vec{h}^I - h^I_3)}, \nonumber \\
    F^I_+ &= -\frac{1}{2|\vec{h}^I|^3} \vec{h}^I \cdot \left(\pdv{\vec{h}^I}{k_y} \times \pdv{\vec{h}^I}{k_z}\right).
\end{align}
Eq.\ \eqref{eq:ysym-plane1} implies that $h_\mu^I$ is even in $k_y$. Therefore, $F_+^I$ is odd in $k_y$, so the integral of $F^I_+$ vanishes on plane I. Moreover, per Eq.\ \eqref{eq:ysym-plane1}, $(\vb{A}^I_+)^z$ is even in $k_y$, so $(\vb{A}^I_+)^z(\pi, k_z) = (\vb{A}^I_+)^z(-\pi, k_z)$. Thus,
\be
n_I = 0.
\ee

\subsection{Plane II Contribution} \label{sec:planeII}
The $n_{II}$ integer is similarly given by 
\begin{align*}
    2\pi n_{II} = \int_{I} \tr F^{II}_+ - \lambda^{II}_+(\pi) + \lambda^{II}_+(-\pi).
\end{align*}

We simplify the above expression by relating the two Berry connection terms with the $S$ screw symmetry (because $S \cdot (\pi, \pi, k_z) = (\pi, -\pi, k_z)$). Then, for $\vb{k} = (\pi, \pi, k_z)$, $H^{II}(S\vb{k})$ is given by 
$$B_X(S\vb{k})^\dagger M_S(\vb{k}) B_X(\vb{k}) H^{II}(\vb{k}) B_X(\vb{k})^\dagger M_S(\vb{k})^\dagger B_X(S\vb{k}).$$
The effects of this matrix transformation are best visualized in cylindrical coordinates, i.e., $(h_1, h_2) = h_r(\cos \phi, \sin \phi)$. Under the matrix transformation,
\begin{align}\label{eq:screw_relations_planeII}
    h_3^{II}(\pi, k_z) &= - h_3^{II}(-\pi, k_z) \nonumber \\
    h_r^{II}(\pi, k_z) &= h_r^{II}(-\pi, k_z)\nonumber \\
    -\phi^{II}(\pi, k_z) &= \phi^{II}(-\pi, k_z) - k_z + \pi.
\end{align}
In cylindrical coordinates, the Berry connection is 
\be
\vb{A}^{II}_+ = \frac{|\vec{h}^{II}| + h_3^{II}}{2 |\vec{h}^{II}|} \grad_{\vb{k}} \phi, \nonumber
\ee
and thus $\vb{A}^{II}_+(\pi, k_z)$ and $\vb{A}^{II}_+(-\pi, k_z)$ are related by
\begin{align}\label{eq:screw_planeII}
    \vb{A}^{II}(\pi, k_z) =\vb{A}^{II}(-\pi, k_z) + \left.\grad_{\vb{k}} \phi^{II}\right|_{(\pi, k_z)} + \nonumber\\
    \frac{1}{2}\left(-1 + \frac{h_3^{II}(\pi, k_z)}{|\vec{h}^{II}|(\pi, k_z)}\right) \vu{z}.
\end{align}
Therefore, 
\begin{align}\label{eq:nII_calculation}
    2 \pi n_{II} = \int_{-\pi}^\pi \int_{-\pi}^\pi F^{II}_+(k_y, k_z) \dd k_y \dd k_z  -\Delta \phi^{II}(\pi) + \nonumber \\
    \pi - \int_{-\pi}^\pi \frac{h_3^{II}(\pi, k_z)}{2|\vec{h}^{II}|(\pi, k_z)} \dd k_z,
\end{align}
where 
\be \label{eq:phi_definition}
\Delta \phi^{II}(k_y) = \int_{-\pi}^{\pi} \left.\pdv{\phi^{II}}{k_z}\right|_{(k_y, k_z)} \dd k_z.
\ee

\subsubsection{Plane 0 Contribution}
Recall that 
\begin{align*}
    2\pi n_{0} = &\int_{0} \tr F - \lambda^{II}_+(\pi) - \lambda^{II}_+(-\pi) +\lambda^I_+(\pi) + \lambda^I_+(-\pi).
\end{align*}
Using Eq.\ \eqref{eq:screw_planeII}, we find that the Berry connection contribution from plane II, $-\lambda_+^{II}(-\pi) - \lambda_+^{II}(\pi)$, is given by 
\begin{align} \label{eq:n0_planeII_contribution}
-2\lambda_+^{II}(-\pi) -\Delta \phi^{II}(\pi) +
\pi - \int_{-\pi}^\pi \frac{h_3^{II}(\pi, k_z)}{2h^{II}(\pi, k_z)} \dd k_z. 
\end{align}
We likewise compute the plane I contribution using that $S \cdot (0, \pi, k_z) = (\pi, 0, k_z)$. Then, for $\vb{k} = (0, \pi, k_z)$, $H^{II}(S\vb{k})$ is given by
$$B_X(S\vb{k})^\dagger M_S(\vb{k}) B_X(\vb{k}) H^{I}(\vb{k}) B_X(\vb{k})^\dagger M_S(\vb{k})^\dagger B_X(S\vb{k}).$$
This transformation gives us
\begin{align*}
    -h_3^{II}(0, k_z) &= h_3^I(\pi, k_z),  \\
    h_r^{II}(0,k_z) &=  h_r^{I}(\pi,k_z), \\
    -\phi^{II}(0, k_z) + k_z - \pi &= \phi^I(\pi, k_z).
\end{align*}
Therefore, the Berry connection contribution from plane I is given by 
\begin{align}\label{eq:n0_planeI_contribution}
    2\lambda_+^{II}(0) -2\Delta \phi^{II}(0) + 2\pi - \int_{-\pi}^\pi \frac{h_3^{II}(0, k_z)}{h^{II}(0, k_z)} \dd k_z. \end{align}
Finally, we compute the Berry curvature contribution from half-plane 0. Because half-plane 0 is invariant under the $Y$-glide, the $Y$-eigenbasis is a convenient choice of basis for computing the Berry curvature integral. As with the $X$ glide, the $Y$ glide splits the filled states of $H_{106}$ into a positive and negative sector on half-plane 0. Therefore,
\begin{align*}
\int_0 \tr F = \int_0 F^0_+ + F^0_-.
\end{align*}
We also compute this integral by equating it to an integral on plane II, this time using the product of symmetries $YS$, which maps $(k_x, \pi, k_z)$ to $(\pi, k_x, k_z)$. The $YS$ symmetry transformation gives us 
\begin{align*}
    h_3^0(k_x, k_z) &= h_3^{II}(k_x + 2\pi, k_z)\\
    h_r^0(k_x, k_z) &= h_r^{II}(k_x + 2\pi, k_z)\\
    \phi^0(k_x, k_z)&= \phi^0(k_x + 2\pi, k_z) - k_z.
\end{align*}
Under this transformation, the Berry connections on half-plane 0 and plane II have the following relation:
\begin{align*}
\vb{A}^0_+(k_x, k_z) =  &- \left(1 + \frac{h_3^{II}(k_x + 2\pi, k_z)}{2|\vec{h}^{II}(k_x + 2\pi, k_z)|}\right) \vu{z} + \\
& \vb{A}^{II}_+(k_x + 2\pi, k_z)
\end{align*}
We thus get a boundary term for the Berry curvature:
\begin{align*}
F^0_+(k_x, k_z) =  &- \partial_{k_x}\left(1 + \frac{h_3^{II}(k_x + 2\pi, k_z)}{2|\vec{h}^{II}(k_x + 2\pi, k_z)|}\right) + \\
& F^{II}_+(k_x + 2\pi, k_z).
\end{align*}

Because $\left(1 + \frac{h_3^{II}}{2|\vec{h}^{II}|}\right)$ is an analytic function on the Brillouin zone, we can apply Stokes' theorem to the boundary term. From Eqs.\ \eqref{eq:ysym-plane1} and \eqref{eq:screw_relations_planeII} and from the $4\pi$ periodicity in $k_y$,
\begin{align*}
    h_3^{II}(0, k_z) &= h_3^{II}(2\pi, k_z)\\
    h_3^{II}(\pi, k_z) &= -h_3^{II}(3\pi, k_z).
\end{align*}
Then, because $F^{II}_+(k_y + 2\pi, k_z) = F^{II}_-(k_y, k_z)$,
\begin{align}\label{eq:n0_plane0_contribution}
    \int_0 \tr F = &\int_{-\pi}^\pi \dd k_z \int_{0}^\pi \dd k_y (F^{II}_+ + F^{II}_-) +  \nonumber\\
    &\int_{-\pi}^\pi \dd k_z \frac{h_3^{II}(0,k_z)}{|\vec{h}^{II}(0,k_z)|}.
\end{align}

Combining the three contributions to $n_0$ gives us 
\begin{align}
2\pi n_0 = \int_{-\pi}^\pi \dd k_z \int_0^\pi \dd k_y \text{ } (F_+^{II} + F_-^{II}) \dd k_y \dd k_z + \nonumber \\
2\lambda_+^{II}(0) - 2\lambda_+^{II}(-\pi) + 
3\pi - \nonumber \\ \int_{-\pi}^\pi \frac{h_3^{II}(\pi, k_z)}{2h^{II}(\pi, k_z)} - \Delta \phi^{II}(\pi) - 2 \Delta \phi^{II}(0).
 \label{eq:n0_calculation}
\end{align}

\subsection{Total Glide Invariant}
Because $n_I = 0$ and because we are working in $\mathbb{Z}_2$, the total glide invariant is given by $ n =n_0 - n_{II}$. To combine Eqs.\ \eqref{eq:nII_calculation} and \eqref{eq:n0_calculation}, we first apply 
$$\int_{-\pi}^\pi \int_0^\pi F_-^{II} \dd k_y \dd k_z = -\int_{-\pi}^\pi \int_{-\pi}^0 F_+^{II} \dd k_y \dd k_z$$
to our expression for $n_{II}$. Then, we find that  
\begin{align}\label{eq:glide_invariant}
    2\pi n=  & -2\left[\int_{-\pi}^\pi \int_{-\pi}^0 F_+^{II} \dd k_y \dd k_z - \lambda_+^{II}(0) + \lambda_+^{II}(-\pi)\right] +\nonumber \\ 
    &
    2\pi  - 2 \Delta \phi^{II}(0).
\end{align}

The bracketed term in Eq.\ \eqref{eq:glide_invariant} is the difference between a Berry connection integral on a closed loop and the Berry curvature integral within the closed loop. Thus, when multiplied by $-2$, this term is an integer multiple of $4\pi$. Additionally, according to Eq.\ \eqref{eq:phi_definition}, $\Delta \phi^{II}(0)$ is the change in phase of the function $h^{II}_1(0, k_z) - i h^{II}_2(0, k_z)$ over the loop $k_z = -\pi \to k_z = \pi$. If $h^{II}_1 + i h^{II}_2$ is always nonzero, this change in phase is always a multiple of $2\pi$, and if $h^{II}_1 + ih^{II}_2$ is $0$ at some $(0,k_z)$, we can add a small symmetry-preserving perturbation to move the zero outside of the loop $(0, k_z)$. Therefore, the last term in Eq.\ \eqref{eq:glide_invariant}, $2\Delta \phi^{II}(0)$, is also an integer multiple of $4\pi$. Thus, after taking Eq.\ \eqref{eq:glide_invariant} modulo $4\pi$, only the $2\pi$ remains, and so the glide invariant $n =1\mod 2$, i.e., it is nontrivial. 

\section{Discussion} \label{sec:discussion}
We have established that a $4$ band, filling $2$ feQBI realized by fermions in SG 106 and the symmetry class A (i.e., spinless fermions without time-reversal symmetry) has a nontrivial magnetoelectric polarizability and thus is an axion insulator. Can this result be strengthened to show that all insulators in SG 106 at the non-atomic filling of $4n+2$ are axion insulators with axion angle $\theta = \pi$, while all insulators at filling $4n$ have the trivial axion angle $\theta = 0$? If this stronger statement were valid, SG 106 would display a one-to-one correspondence between filling and topological invariant previously exhibited in 2D crystals with external magnetic fields \cite{Lu_Ran_Oshikawa_2020}. Although we only conjecture the validity of this stronger statement instead of providing a complete argument, the following observations explain the rationale behind our conjecture. 

First, we observe that the non-atomic filling cannot be realized by a weak phase, which in our setting corresponds to a stack of Chern insulators in real space (the two orthogonal glides forbid such a phase anyway, as we argue in Appendix \ref{appendix-chern}). If such a weak phase were consistent with the non-atomic filling, then we could go to the decoupled limit, and each of the decoupled planes would also carry a non-atomic filling. Because symmetry analysis dictates that no feQBI is possible in two dimensions for class A \cite{106feQBI}, we conclude the feQBI must correspond to a 3D phase without any decoupled plane limit.

Now, suppose the strong factor is either the group $\mathbb Z$ or $\mathbb Z_m$, where $m$ denotes a positive integer. In both cases, there exists a reference non-trivial insulator that generates all the entries in the strong factor through the following operation: to generate the entry indexed by $n$, we first take $n$ copies of the reference state. Next, we allow for arbitrary symmetry- and gap-preserving coupling between the different copies while allowing for the addition of any trivial atomic insulator with the same set of symmetries.
Now, if the axion angle of the reference ``generating insulator'' were $0$, then all the entries in the strong factor would also have $\theta = 0$. Because the nontrivial phase we have found in this paper would not be compatible with any of the entries in the classification, this would be impossible. The same argument applies to the non-atomic filling, i.e., if the generating insulator had an atomic filling of $4n$, then again all band insulators would have filling $4n$. As such, we have established that the feQBIs here always have a nontrivial axion angle, which leads to our conjecture.

The argument above relies on the assumption that the strong factor of the (stable) classification has a single generator. 
We leave this as a conjecture, but remark that it is a plausible one given that the same property holds for the corresponding discussion of topological (crystalline) insulators in class AII \cite{Song2018, PhysRevX.8.031070, Song2019}.
To complete the argument, one can adopt the frameworks in Refs.~\onlinecite{PhysRevX.7.011020, PhysRevX.8.011040, PhysRevB.96.205106, Songeaax2007, PhysRevB.99.115116} for the classification of topological crystalline phases and derive the full classification, which could then be compared with the $\mathbb Z_2$ axion angle. We also note that, in fact, the Hamiltonian we introduced in Sec.\ \ref{sec:model_feqbi} 
could also be understood in the mentioned classification frameworks.
Alternatively, it might be possible to establish the conjecture by a systematic investigation on the possible Wilson loop behavior for any filling $4n+2$ insulator.

Lastly, we point out that the filling-enforced nature of the nontrivial phases at filling $4n+2$ would imply that it is impossible to drive a symmetry-respecting transition from the nontrivial phase to a trivial phase \cite{feqbi-soc}.
In other words, the nontrivial phase considered here cannot be understood through a band-inversion paradigm \cite{colloquium_top_insulator, top_ins_sc, franz_molenkamp_2013}.
More specifically, if, as we have proposed, the notion of feQBIs exactly maps onto the familiar notion of an axion insulator in our symmetry setting (class A in space group 106), the axion insulator here cannot be accessed from a trivial, atomic state through the tuning of the symmetry-allowed mass term of an emergent Dirac fermion. 
Thus, if our conjecture holds, a more detailed understanding on why such a transition is forbidden by the symmetries on hand is an interesting open problem, especially when one incorporates interactions. Alternatively, if the conjecture does not hold, then there must be stable topological crystalline insulators in the present symmetry setting which have an axion angle of $\theta = 0\mod 2\pi$ and a filling of $4n+2$. Given that the spatial symmetries involved are only glides, screws, and a two-fold rotation, we are not aware of a candidate state and as such this alternative scenario will also be interesting in its own right.

\begin{acknowledgments}
The work of AK was supported by the National Science Foundation Graduate Research Fellowship under Grant No.\ 1745302. AK also acknowledges support from the Paul and Daisy Soros Fellowship and the Barry M.\ Goldwater Scholarship Foundation. AV was supported by the Center for the Advancement of Topological Semimetals, an Energy Frontier Research Center funded by the U.S. Department of Energy Office of Science, Office of Basic Energy Sciences through the Ames Laboratory under its Contract No. DE-AC02-07CH11358. The work of HCP was supported by a Pappalardo Fellowship at MIT and a Croucher Foundation Fellowship. We would like to thank an anonymous referee for their insights on the extension of our argument for all feQBIs in SG 106 via a Wilson-loop-based argument. 
\end{acknowledgments}

\appendix
\section{Tight-Binding feQBI Model}\label{appendix-full-hamiltonian}
Here, we present the result of applying Eq.\ \eqref{eq:sym_ansatz} to Eq.\ \eqref{eq:ansatz}. We provide the upper triangular block of the Hamiltonian below:
\begin{widetext}
\begin{align}\label{eq:full_model_hamiltonian}
H_{11} = H_{22} = \frac{t_x}{2} \cos(k_x), \quad H_{33} =  H_{44} = \frac{t_x}{2} \cos(k_y), \quad H_{23} = \left[\frac{-t_d}{8} \left(e^{-ik_x} + e^{ik_y}\right) - \frac{t_d^\star}{8} e^{-ik_z} \left(1 + e^{i(k_y - k_x)}\right)\right], \nonumber\\
H_{12} =  H_{34} = 0,  \quad 
H_{13} = \frac{1}{4}\left[t_z + t_c \cos(k_x - k_y)- e^{-ik_z}\left(t_z^\star + t_c^\star \cos(k_x - k_y)\right) \right], \nonumber\\ H_{24} = \frac{1}{4}\left[e^{-ik_z}\left(t_z^\star + t_c^\star \cos(k_x + k_y)\right) -t_z - t_c \cos(k_x + k_y)\right], \quad H_{14} = \left[\frac{t_d}{8} \left(1 + e^{i(k_x - k_y)}\right) + \frac{t_d^\star}{8} e^{-ik_z} \left(e^{-i k_y} + e^{ik_x}\right)\right].
\end{align}
\end{widetext}

\section{Berry Connection after X Glide Basis Transformation} \label{appendix-glide}
Here, we derive the form of the Berry connection for a filled state written as a linear combination of X Glide eigenstates. Recall that the X glide eigenbasis is
$$
B_X(\vb{k}) = \frac{1}{\sqrt{2}}\begin{pmatrix}e^{-ik_y/2} & 0 & -e^{-ik_y/2} & 0 \\
1 & 0 & 1 & 0 \\
0 & -e^{-ik_y/2} & 0 & e^{-ik_y/2}\\
0 & 1 & 0 & 1
\end{pmatrix}.
$$
Let $b_1^+(\vb{k})$ and $b_2^+(\vb{k})$ be the first two columns of $B_X(\vb{k})$, and let $b_1^-(\vb{k})$ and $b_2^-(\vb{k})$ be the last two columns. Then, on the glide invariant planes $k_x = 0$ or $k_x = \pi$, we can write any positive sector filled state of the SG 106 Hamiltonian as 
$$
v^+(\vb{k}) = a_1^+(\vb{k}) b_1^+(\vb{k}) + a_2^+(\vb{k}) b_2^+(\vb{k}).
$$
In other words, if $v^+$ is a filled eigenstate of the original SG 106 Hamiltonian, $(a_1^+, a_2^+, 0, 0)$ is a filled eigenstate of $H^p$ in Equation  \eqref{block-diagonal-hamiltonian}. 

We now compute the Berry connection. Let $$\vb{A}^p_+ = -i v^+(\vb{k})^\dagger \grad_{\vb{k}} v^+(\vb{k})$$
be the true Berry connection on the glide invariant plane $p$, and let $$\tilde{\vb{A}}^p_+ = - i \sum_j a_j^+(\vb{k})^\dagger \grad_{\vb{k}} a_j^+(\vb{k})$$
be the Berry connection computed directly from the eigenstates of $H^p$. 
We expand the true Berry connection $\vb{A}^p_+$ as follows:
$${\vb{A}}^p_+ = -i \sum_{j,l} a_j^+(\vb{k})^\star b_j^+(\vb{k})^\dagger \grad_{\vb{k}}\left(a_l^+(\vb{k}) b_l^+(\vb{k})\right).$$
From the form of the $X$ glide eigenvectors, we have that $b_j^+(\vb{k})^\dagger b_l(\vb{k}) = \delta_{jl}$ and $b_j^+(\vb{k})^\dagger \grad_{\vb{k}}b_l^+(\vb{k}) = \delta_{jl} b_j^+(\vb{k})^\dagger \grad_{\vb{k}}b_j^+(\vb{k})$. Then,
\begin{align*}
    \vb{A}^p_+ = -i \sum_{j}  \left|a_j^+(\vb{k})\right|^2 b_j^+(\vb{k})^\dagger \grad_{\vb{k}}b_j^+(\vb{k}) +\\
    a_j^+(\vb{k})^\star  \grad_{\vb{k}}\left(a_j^+(\vb{k})\right) .
\end{align*}
We compute the first term explicitly using our expression for $B_X(\vb{k})$. The second term is just $\tilde{\vb{A}}^p_+$. We thus get 
$$\vb{A}^p_+ = \tilde{\vb{A}}^p_+ + (-1/2, 0).$$
Therefore, $(\vb{A}^p_+)^z = (\tilde{\vb{A}}^p_+)^z$ and $F^P_+ = \tilde{F}^P_+$, or in other words, we can just use the  eigenstates of $H^p$ in the process of computing the glide invariant. 

\section{Zero Net Chern Number on Brillouin Zone Planes}\label{appendix-chern}

In this section, we argue that for any general SG 106 insulator, the net Chern number arising from the filled bands on any Brillouin zone plane (constant $k_x$, constant $k_y$, or constant $k_z$) is $0$. Note that either the $X$ or the $Y$ symmetry will map a Brillouin zone plane to itself. Without loss of generality, we restrict our attention to the constant $k_z$ plane $k_z = k_z^0$. Let us take the point $(k_x^0, k_y^0)$ and restrict our attention to an open neighborhood around this point and the corresponding open neighborhood around $(-k_x^0, k_y^0)$ from the glide. Suppose we have $n$ total bands, and $\ell$ filled bands. Let the filled states of the insulator in these open neighborhoods be given by the $n \times \ell$ matrix $B(k_x, k_y)$, where each column corresponds to a filled eigenstate. Note that there exists a locally smooth $\ell \times \ell$ unitary matrix function $U(k_x, k_y)$, also known as the sewing matrix, such that
\begin{equation} \label{eq:sewing_matrix}
    M_X(k_x, k_y) B(k_x, k_y) U(k_x, k_y) = B(-k_x, k_y).
\end{equation}
Here, $X$ is the $n\times n$ glide symmetry matrix . 
The trace of the Berry curvature at $(k_x^0, k_y^0)$ is given by \begin{align*}
    \tr F(k_x^0, k_y^0) = & \tr i\left.\pdv{B^\dagger}{k_y}\right|_{(k_x^0, k_y^0)}\left.\pdv{B}{k_x}\right|_{(k_x^0, k_y^0)}- \\
   & \tr i\left.\pdv{B^\dagger}{k_x}\right|_{(k_x^0, k_y^0)}\left.\pdv{B}{k_y}\right|_{(k_x^0, k_y^0)}.
\end{align*}
We compute the Berry curvature at $(-k_x^0, k_y^0)$ using
\begin{align*}
     \left.\pdv{k_x}\left(X(-k_x,k_y) B(-k_x, k_y) U(-k_x,k_y)\right)\right|_{(-k_x^0, k_y^0)} = \\
     -\left.\pdv{k_x}\left(X(k_x,k_y) B(k_x, k_y) U(k_x,k_y)\right)\right|_{(k_x^0, k_y^0)}.
\end{align*}
Then, the trace of the Berry curvature at $(-k_x^0, k_y^0)$, assuming that all derivatives are taken at $(k_x^0, k_y^0)$, is
\begin{widetext}
\begin{align}
\tr F(-k_x^0, k_y^0) =  &i\tr \left( \partial_{k_x}\left(U^\dagger B^\dagger X^\dagger\right)\partial_{k_y}\left(XBU\right) - \partial_{k_y}\left(U^\dagger B^\dagger X^\dagger\right)\partial_{k_x}\left(XBU\right) \right) \nonumber \\
= & i \tr \left(\left(\partial_{k_x} U^\dagger\right) \left(\partial_{k_y} U\right)  - \left(\partial_{k_y} U^\dagger\right) \left(\partial_{k_x} U\right) \right)  +
i \tr \left(\left(\partial_{k_x} X^\dagger\right) \left(\partial_{k_y} X\right)  - \left(\partial_{k_y} X^\dagger\right) \left(\partial_{k_x} X\right) \right)  - \tr F(k_x^0, k_y^0) + \nonumber \\
 & i \tr \left((\partial_{k_x} B^\dagger)X^\dagger (\partial_{k_y} X) B - B^\dagger (\partial_{k_y}X^\dagger) X (\partial_{k_x}B) + B^\dagger(\partial_{k_x} X^\dagger) X(\partial_{k_y} B) -  (\partial_{k_y}B^\dagger) X^\dagger (\partial_{k_x}X)B\right) + \nonumber \\
 & i \tr \left((\partial_{k_x} U^\dagger)(XB)^\dagger (\partial_{k_y} (XB)) U - U^\dagger (\partial_{k_y}(XB)^\dagger) (XB) (\partial_{k_x}U)\right) + \nonumber \\
 & i \tr \left(U^\dagger(\partial_{k_x} (XB)^\dagger) XB(\partial_{k_y} U) -  (\partial_{k_y}U^\dagger) (XB)^\dagger (\partial_{k_x}(XB))U\right) .
 \label{eq:sewingmatrixcurvature}
\end{align}
\end{widetext}
Using the cyclic property of the trace and that $\partial(A^\dagger A) = 0$ for any matrix $A$, we find that the last three lines of Eq.\ \eqref{eq:sewingmatrixcurvature} go to $0$. Thus, we are left with 
\begin{align}\label{eq:reduced_sewing_matrix_curvature}
    i \tr \left(\left(\partial_{k_x} U^\dagger\right) \left(\partial_{k_y} U\right)  - \left(\partial_{k_y} U^\dagger\right) \left(\partial_{k_x} U\right) \right)  + \nonumber \\
    i \tr \left(\left(\partial_{k_x} X^\dagger\right) \left(\partial_{k_y} X\right)  - \left(\partial_{k_y} X^\dagger\right) \left(\partial_{k_x} X\right) \right)  + \nonumber \\ -
    \tr F (k_x^0, k_y^0).
\end{align}
Let us now consider the first line of Eq.\ \eqref{eq:reduced_sewing_matrix_curvature}. This term is effectively the trace of the Berry curvature for a insulator with filled states given by the orthonormal columns of $U$. The columns of $U$ form a complete basis, so all bands of this hypothetical insulator are filled, and thus the trace of the Berry curvature for this insulator is $0$. By the same argument, the second term is $0$, so
$$\tr F(-k_x^0, k_y^0) = \tr F(k_x^0, k_y^0),$$
and thus the Chern number is given by
$$c =  \int_{k_z = k_z^0} \tr F = 0.$$

\bibliography{sg106.bib}

\end{document}